# Extreme Variability of the V488 Persei Debris Disk

Rieke, G. H.,[1,2] Su, K. Y. L.,[1] Melis, Carl,[3] and Gáspár, András[1]

[1]*Department of Astronomy/Steward Observatory, The University of Arizona, Tucson, AZ 85721-0009, USA*
[2]*Department of Planetary Sciences/Lunar & Planetary Laboratory, The University of Arizona, 1629 E University Blvd, Tucson AZ 85721-0092, USA*
[3]*Center for Astrophysics and Space Sciences, University of California, San Diego, CA 92093-0424, USA*

## ABSTRACT

V488 Persei is the most extreme debris disk known in terms of the fraction of the stellar luminosity it intercepts and reradiates. The infrared output of its disk is extremely variable, similar in this respect to the most variable disk known previously, that around ID8 in NGC 2547. We show that the variations are likely to be due to collisions of large planetesimals ($\geq$ 100 km in diameter) in a belt being stirred gravitationally by a planetary or low-mass-brown-dwarf member of a planetary system around the star. The dust being produced by the resulting collisions is falling into the star due to drag by the stellar wind. The indicated planetesimal destruction rate is so high that it is unlikely that the current level of activity can persist for much longer than $\sim$ 1000 - 10,000 years, and it may signal a major realignment of the configuration of the planetary system.

## 1. INTRODUCTION

Planetary debris disks reveal critical stages in planetary system evolution (Wyatt 2008; Gáspár et al. 2013). An important development is finding debris disk behavior tracing recent major collisions between large planetesimals. These characteristics include either extreme infrared excesses (Gorlova et al. 2007; Balog et al. 2009; Melis et al. 2010) or spectral features indicative of finely divided silica particles possibly associated with condensation out of vapor from very violent collisions (e.g. Rhee et al. 2008; Johnson et al. 2012). Many of these cases are too far in their evolution to be associated with the protoplanetary disk phase ($\leq$ 10 Myr), but represent the later stages of planetary system formation and evolution when, for example, terrestrial planets are growing (Meng et al. 2017). Understanding these systems can illuminate a critical stage in the building of such planets in the first 100 − 200 Myr of evolution of a planetary system. In addition, some of these systems are beyond this age range (Moór et al. 2021; Melis et al. 2021) and indicate that the phenomenon may also result from major shifts in the structure of exoplanetary systems

It was expected that the aftermaths of giant planetesimal collisions would evolve slowly on a human scale, e.g. over millions of years (e.g. Grogan et al. 2001; Booth et al. 2009). Surprisingly, some debris disks are variable over a few years, or even faster (Melis et al. 2012; Meng et al. 2012, 2015). These systems stretch our understanding but can be explained through specific conditions hypothesized to result from the smashup of bodies at least the size of large asteroids (Meng et al. 2014; Su et al. 2019). Indeed, we now have direct evidence for this process in the case of Fomalhaut (Gaspar & Rieke 2020). An extreme example of this behavior, and thus the most pressing for development of physical models, has been the system ID8 in NGC 2547, where substantial changes in the disk output are seen on monthly timescales (Meng et al. 2014). In this paper, we report on a second system that is nearly as rapidly variable as ID8, demonstrating that the rapid brightness fluctuations of ID8 are not unique. This disk orbits V488 Per, an ~ 80 Myr old K2-2.5V star at a distance of 173 pc in the $\alpha$ Persei Cluster.

V488 Per = AP 70 = WEBDA 1570 was originally identified as a member of the $\alpha$ Persei Clusterby Stauffer et al. (1985), who also showed it to be a BY-Draconis variable (i.e., showing periodic variations associated with starspots). Its properties were further explored by Prosser (1992); Randich et al. (1996); Mermilliod et al. (2008); Cantat-Gaudin et al. (2018). The last of these references utilized Gaia data to establish cluster membership at very high probability. Zuckerman et al. (2012) reported its huge mid-infrared excess, which they describe as "16% (or more) of the stellar luminosity;this is a larger excess fraction than that of any other known dusty main-sequence star." The age ofthe star based on its membership in the $\alpha$ Persei Cluster is 80 Myr (Soderblom et al. 2014), far beyond the age for a residual protoplanetary disk (Meng et al. 2017), so this phenomenon must be linked to secondary dust production in a collision between planetesimals in a young planetary systemaround the star.

In addition to its significance in confirming the extreme form debris disk variability can take, V488 Per is of unique interest for two reasons: (1) the star is of later spectral type and lower luminosity than previously known examples of highly variable extreme disks, so the processes occurring aroundit may be driven by additional physics; and (2) at 80 Myr, the star is older than many other examples of highly variable disks. In this paper, we explore its behavior in light of these characteristics. Wefirst present the relevant observations (Section 2), then analyze them to derive the behavior of thedisk and the possible underlying drivers for the activity (Section 3), and finally summarize our results(Section 4).

## 2. OBSERVATIONS

### 2.1. *Astrometry*

Given the exceptional characteristics of V488 Per interpreted as a debris disk, we have tested whether it could be a chance alignment of the star with a (very peculiar) background source. We show positions of the star (GAIA and 2MASS) and of the excess (WISE and Herschel/PACS (corrected for proper motion)) in Table 1; they coincide virtually perfectly, i.e., within ~ 0".1 for the three at shorter wavelengths and within 0".5 for PACS at 70 $\mu$m. In the latter case, the beam is ~ 5" in diameter and the ratio of signal to noise only 14:1, so the standard indicator of positional accuracy of beamwidth divided by signal to noise would be 0".35, a bit larger than the archive estimates. We show the resulting more conservative errors in the table. To put a probability on this circumstance, we searched the WISE catalog for sources within 1000" of V488 Per, as bright or brighter than it is, and with a W1 - W2 color > 1 mag (to match its color). There was only one such source, indicating that the probability that the excess of the star results from a chance juxtaposition is $\leq 10^{-6}$, assuming a matching radius of 1". We do not consider this possibility further.

## 2.2. Light Curves

We now discuss observations in the infrared and optical that document the variability of V488 Per and its debris disk. There are two sets of infrared data: one from *Spitzer*/IRAC and the other from WISE as described in Section 2.2.1, and three sets of optical data as described in Section 2.2.2.

Table 1. Wavelength-dependent positions for V488 Per

| Source | RA | DEC |
|---|---|---|
| GAIA EDR3 | 03 28 18.683 | 48 39 48.192 |
| 2MASS | 03 28 18.683 | 48 39 48.222 |
| ALLWISE | 03 28 18.699 | 48 39 47.97 |
| PACS[a] | 03 28 18.63 ± 0.04 | 48 39 47.8 ± 0.35 |

[a]position from PACS Point Source Catalog at 70 $\mu$m (Marton et al. 2017)

### 2.2.1. Infrared light curve

During the *Spitzer* warm mission, we monitored V488 Per regularly to search and characterize its infrared viability. V488 Per had two ~ 45-day visibility windows every year for *Spitzer*, separated by ~ 120 days. We report in Figure 1 *Spitzer*/IRAC warm mission observations made under programs 11093, 13014, and 14226 from April, 2015 through the end of the mission in early 2020, providing a total time baseline of ~ 1800 days. All observations used both the 3.6 and 4.5 $\mu$m IRAC wavebands(IRAC1 and IRAC2). Sampling frequencies were typically once per three days. We used a frametime of 30 seconds with 10 cycling dithering positions for both bands, achieving a typical signal-to-noise ratio of 150-300. The dithering pattern was designed to use a number of random pixelpositions to average the intrapixel sensitivity variations of the detector[1]. Furthermore, V488 Per was observed once with IRAC (AOR 18853632, PID: 30717) during the *Spitzer* cold mission as reportedby Zuckerman et al. (2012). For consistency, we also conducted our own photometry on these data.

The data were processed by the *Spitzer* Science Center (SSC) with IRAC pipeline S18.18.0 for the cold mission and with S19.2.0 for the warm mission. We used the BCD (basic calibrated data) images,which have a native scale of 1".22 pixel$^{-1}$. We performed aperture photometry using an on-source radius of 3 pixels and a sky annulus between 12 and 20 pixels, with aperture correction factors of 1.112 and 1.113 for 3.6 and 4.5 $\mu$m, respectively. The BCD photometry was corrected for the pixel solid angle (i.e., distortion) effects based on the measured target positions using files provided

---

[1] Spitzer Science Center 2013, Memo to Spitzer observers regarding high precision photometry using dithered observations; http://ssc.spitzer.caltech.edu/warmmission/news/18jul2013memo.pdf

by the SSC. We also discarded any photometry when the target was too close to the edge of the detector array, and obtained weighted average photometry for each of the astronomical observation requests (AORs) by rejecting the highest and lowest photometry points in the same AOR. The same procedures were also conducted on the mosaic post-BCD products for comparison. A few photometry points from the post-BCD products were significantly fainter (up to 10%) than the

Table 2. *Spitzer* I1 and I2 Photometry of V488 Per

| AORKey | BMJD_I1 | F_I1 (mJy) | err_I1 (mJy) | BMJD_I2 | F_I2 (mJy) | err_I2 (mJy) |
|---|---|---|---|---|---|---|
| 18853632 | 54004.270160 | 62.00 | 0.25 | 54004.267730 | 56.12 | 4.16 |
| 53435392 | 57129.471620 | 41.81 | 0.25 | 57129.469200 | 46.00 | 0.15 |
| 53435136 | 57132.749360 | 41.71 | 0.32 | 57132.746960 | 46.33 | 0.11 |
| 53434880 | 57135.692400 | 41.58 | 0.26 | 57135.690000 | 46.56 | 0.10 |
| 53434368 | 57138.312940 | 42.39 | 0.21 | 57138.310560 | 46.73 | 0.13 |
| 53433856 | 57141.005870 | 41.68 | 0.26 | 57141.003510 | 46.66 | 0.13 |

Note—Table 2 is published in its entirety in machine-readable format. A portion is shown here for guidance regarding its form and content.

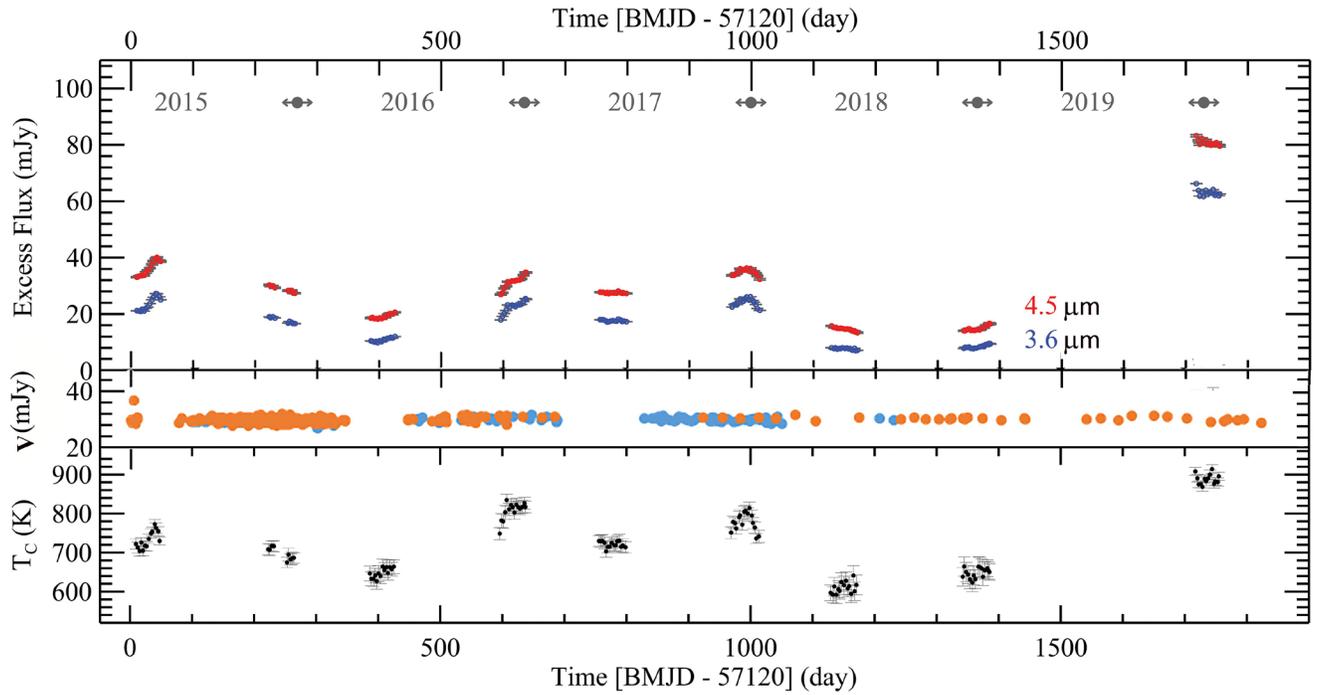

**Figure 1.** Light curves for the entire Spitzer/IRAC monitoring period for V488 Per. Time is given in Julian date along the bottom and in civil date (with double arrows to separate years) along the top. The upper two curves show the evolution of the excess above the stellar photospheric emission at 4.5$\mu$m (red) and 3.6$\mu$m (blue). The middle curve is the optical (V-band) behavior with the zero point suppressed by 20 mJy, shown in greater detail in Figure 3. The lower curve shows the evolution of the color temperature between these two infrared bands, showing that it peaks when the excess emission is also at a peak.

values obtained from the individual BCD images. Inspection of those AORs found that one or two BCD frames have poor WCS (world coordinate system) information, likely causing misalignment in the post-BCD products. Thus we adopted our final photometry based on the weighted average of the BCD images, which were not subject to this issue. Table 2 lists all the IRAC band 1 and band 2 photometry. Figure 1 gives a quick overview for the warm mission data, showing large variations in the infrared excess and its temperature while the V-band brightness of the star stays nearly the same.

The nominal ratio of noise to signal for our measurements, including the star and the disk, at individual epochs is typically 0.3–0.6%. Additional sources of noise typically bring IRAC photometry to the 1% uncertainty level. We have tested our measurements by extracting photometry for six stars within the field for V488 Per, and we find, indeed, that the rms scatter for them is ~1%.

Our fit to the SED of the star (below) suggests that the disk excess emission varies from 30 to 60% of the total signal at 3.6 $\mu$m and from 60–75% at 4.5 $\mu$m, so the nominal errors in the disk signal (after removing the stellar contribution) range from 1–2%. More seriously, any error of the adopted photospheric flux would introduce a systematic error to the disk flux and color, which can hardly be identified in later analysis. As illustrated in Figure 1, the infrared observations show a continuous behavior with no abrupt changes, supporting that the errors are relatively small. Thus, the absolute values of the color temperature of the disk may be biased, but the relative color variations should be more robust. Unlike the case for ID8 (Meng et al. 2014), there is a clear change in the color temperature with output level, with the temperature being higher when the disk is brighter (see lowest panel in Figure 1).

We collected all WISE measurements from NASA/IPAC Infrared Science Archive where single exposure source catalogs are available through different phases of WISE mission (WISE All Sky Database, WISE 3-Band Cryo Database and NEOWISE Reactivation Database). These single expo-sure measurements were averaged over a period of three days, i.e., the same cadence as the *Spitzer* warm mission measurements), rejecting the highest and lowest photometry points. Table 3 and Figure 2 summarize these WISE measurements. They show variations similar in amplitude to those obtained in the *Spitzer* monitoring campaign, but are much less well-sampled. They start well before our more detailed monitoring and show that the excess has been in place at a similar level and degree of variability for at least 15 years. The large increase in excess at the end of our intensive IRAC monitoring series is preceded by about 100 days by a similarly large excess in the WISE bands, and is only beginning to decay a year after it was first detected with WISE. We compared the first epoch of the WISE data shown in Table 3 with the values from the ALLWISE catalog (same data set, but processed differently) and found that the single exposure photometry is ≥ 10% brighter, which is not of concern for this study. The long-term repeatability should be within 3% (Cutri et al. 2013, 2015).

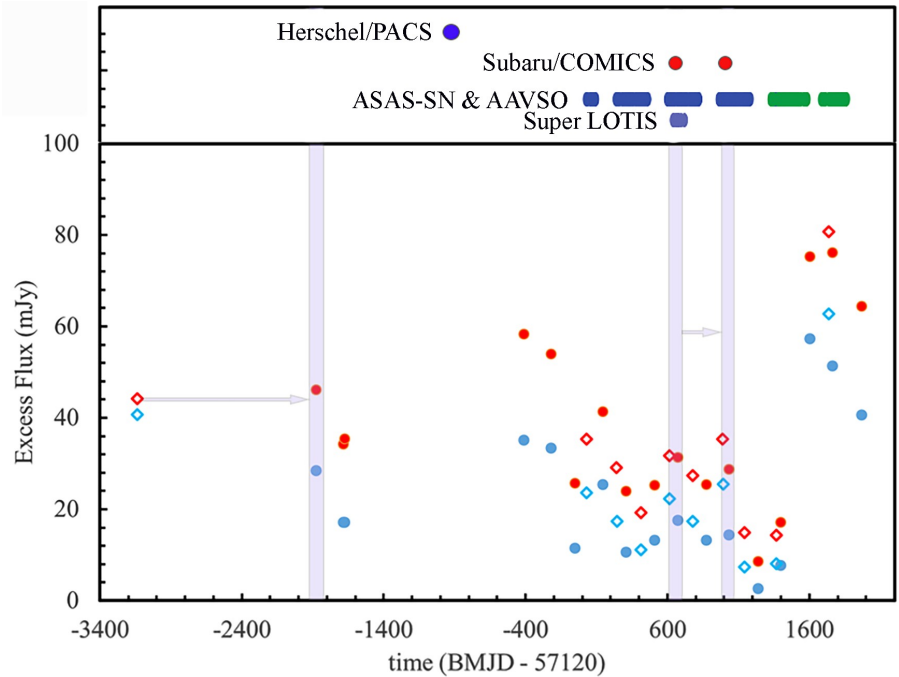

**Figure 2.** Long term infrared light curves from WISE (solid circles) and IRAC (open diamonds) at 3.35or 3.6 $\mu$m (blue) and 4.5 or 4.6$\mu$m (red). The stellar photospheric emission has been subtracted from all the measurements. The earliest two points are from IRAC; the smaller difference between the bands reflects both the smaller wavelength difference between IRAC Bands 1 & 2 than between W1 and W2 and the high color temperature at that time, 1180 200 K . The IRAC measurements at the midpoints of the sampling intervals from Figure 1 are also shown. The first WISE points are from the cryogenic all-sky survey; there are then two from the 3-band survey at the end of the cryogenic mission (identifiable by the double point at 4.6 $\mu$m) and the rest are from NEOWISE and the NEOWISE reactivated missions. Although severely undersampled, these data show that the disk has been highly variable for at least fifteen years, and that the very high value seen in the last set of measurements with IRAC (see Figure 1) was present in the WISE data roughly 100 days earlier - this is a fairly sustained event, but also may well be the largest output overall over more than a decade. The light blue markings illustrate the two models developed in Section 3.5. The upper box shows the timing of the other datasets.

Table 3. WISE W1 & W2 Photometry of V488 Per

| BMJD | W1 (mJy)[a] | W2 (mJy)[a] |
|---|---|---|
| 55243 | 51.7 | 58.7 |
| 55433 | 40.4 | 46.7 |
| 55441 | 40.5 | 47.8 |
| 56707 | 58.7 | 71.1 |
| 56898 | 55.6 | 63.6 |
| 57066 | 34.8 | 38.4 |
| 57260 | 49.0 | 54.0 |
| 57425 | 33.9 | 36.4 |
| 57625 | 36.6 | 37.7 |
| 57792 | 40.9 | 43.8 |
| 57991 | 36.5 | 37.8 |
| 58150 | 37.7 | 41.2 |
| 58355 | 26.0 | 21.0 |
| 58514 | 30.9 | 29.6 |
| 58722 | 80.6 | 87.7 |
| 58878 | 74.6 | 88.6 |
| 59086 | 63.9 | 76.8 |

[a]Errors will be dominated by WISE repeatability, ∼ 3%.

### 2.2.1. *Optical*

V488 Per was identified as a BY-Draconis type variable by Stauffer et al. (1985), i.e., showing modest periodic variations as large starspots rotate into and out of the direction toward the observer. To study the variability of this star, we have used optical V-band photometry from the AAVSO and ASAS-SN (Shappee et al. 2014) archives, plus measurements specifically obtained for this program with Super-LOTIS (Livermore Optical Transient Imaging System), a robotic telescope dedicated to the search for optical counterparts of gamma-ray bursts (GRBs) and located at the Steward Observatory Kitt Peak site. There is a single high value, 35.8 mJy on JD 2457125.57 from the AAVSO data. The fluxes were in the usual range on the previous date and four days later. This could either be a flare, which is common behavior in BY-Dra stars, or a bad measurement. We have omitted it from our further analysis. Figure 3 shows all of the remaining photometry obtained during the same period as the *Spitzer* data.

### 2.3. *Other Observations*

Here, we identify additional measurements that provide additional insights to the V488 Per debris system behavior, but do not have a cadence that reveals the variations of its output.

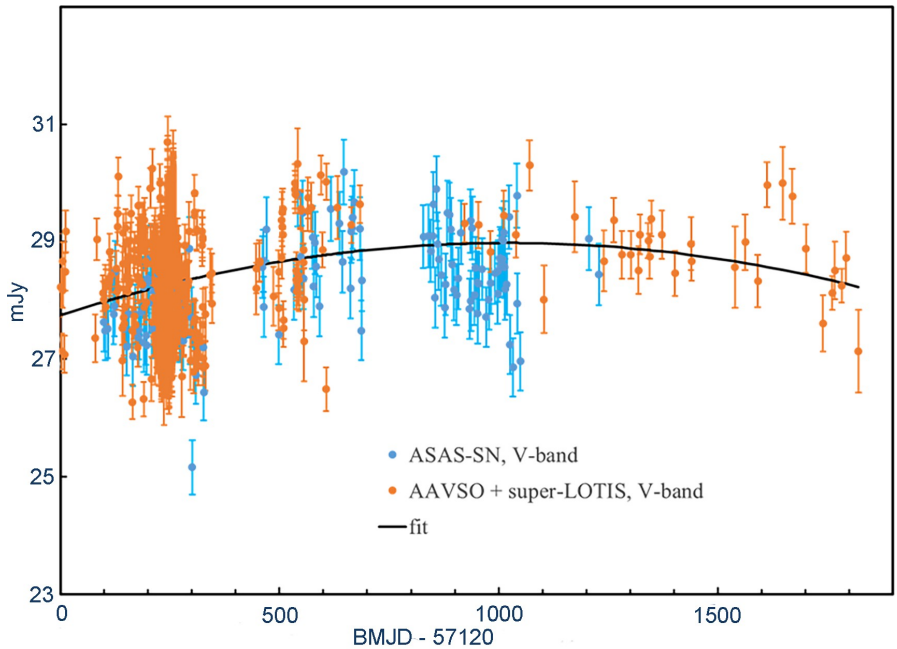

**Figure 3.** Visible light curve for the entire Spitzer monitoring period for V488 Per. The super-LOTIS measurements contribute to the dense set in the first 500 days and are not shown separately. There is a possible weak trend with time, shown by the fit. We also find evidence for a periodic variation with an amplitude of 5.9% for the first year, but < 2% for the remaining time (see text). Removing this effect, the rms scatter is 2.1 - 2.4%.

### 2.3.1. *Additional mid-infrared photometry*

Photometry of V488 Per using intermediate-band filters around 10 $\mu$m was obtained with COMICS (Kataza et al. 2000; Okamoto et al. 2003) on the Subaru Telescope on 14 Jan, 2017 and 28 Jan, 2018 (BMJD - 57120 of 648 and 1027 respectively). The standard chopping by the secondary mirror was used at a frequency of 0.2 Hz; the chopping throw was 15 arcsec. Rather than relying on traditional standard stars and air mass corrections, we observed the nearby star HD 19373, spectral type G0V, and took the WISE measurement of this star to derive the flux calibration. Because the ratio of signal to noise was modest in both runs and the debris disk was of similar brightness (see below), we averaged the measurements for our analysis. These measurements are listed in Table 4 and discussed further in Section 3.5. The flux density for V488 Per observed in the COMICS bands agrees well with the IRAC Band 4 and WISE W3 measurements within the errors, i.e. within the sparse time sampling and limited signal to noise, there is no evidence for variations near 10$\mu$m at amplitudes as large as those seen near 4 $\mu$m.

IRAC photometry was reported by Zuckerman et al. (2012) on BMJD - 57120 ≡ 3115 (Sept. 26, 2006) (AORKEY: 18853632). We have rereduced these measurements with results summarized in Table 5. As shown in Figure 2, these measurements (at the extreme left of the figure) were obtained when the excess was relatively bright.

### 2.3.2. *Herschel*

Herschel PACS 70 and 160 $\mu$m observations were obtained in 2012 Sep 11 (PID OT2_cmelis_3, OBSID 1342250847 and 1342250848) using the PACS mini-scan map mode. We retrieved the archival data from the Herschel Science Center, and reduced the data following the procedure outlined in Balog et al. (2014) to produce the final mosaics. The source is detected at both bands, but with less signal-to-noise in the 160 μm map due to large-scale cirrus present in the data. We used aperture photometry to estimate the PACS flux by fixing the source center at the centroid of the 70 μm position. Aperture sizes of 6″ and 11″ with aperture correction factors of 1.57 and 1.54 were used for the 70 and 160 μm data, respectively. Including the flux calibration uncertainty, the final PACS photometry is 63.3 ± 5.9 mJy in the 70 μm band, and 25.7 ± 12.2 mJy in the 160 μm band. Our 70 μm flux agrees well with the Herschel PACS Point source catalog (Marton et al. 2017): 66.45 ± 6.66 mJy. The fluxes are shown in Table 4.

### 2.3.3. *Radial Velocities*

Table 6 reports measurements of the radial velocity of V488 Per. The weighted average of $0.07 \pm 0.05$ km s$^{-1}$ is consistent with the average for the $a$ Per Cluster ($1.39 \pm 0.17$ km s$^{-1}$ Mermilliod et al. (2008)) within the cluster velocity dispersion of $\leq 1.1$ km s$^{-1}$ (Makarov 2006). The deviation of the individual measurements from the average for the star is insignificant in seven cases ($\leq 2.5\ \sigma$) and marginally significant at 1.73 km s$^{-1}$ for the eighth. The scatter in the APOGEE measurements, which appear to be the most accurate, is $\sim 0.2$ km s$^{-1}$, only slightly larger than the quoted errors[2].

### 2.3.4. *X-ray*

Prosser et al. (1996) report an X-ray detection of V488 Per with the ROSAT Position Sensitive Proportional Counter (PSPC). They find that $\log(L_X$ (ergs/s)$) = 29.68$ and $\log(L_X/L_{bol}) = -3.33$. As a K-dwarf, the detection is expected; the overall detection rate for K-dwarfs in the cluster is 83%, and the X-ray flux from V488 Per is typical for the cluster members (Prosser et al. 1996).

### 3. ANALYSIS

In this section, we first show that V488 Per is a main sequence K2-2.5V star, younger but analogous in many ways to eps Eridani, a similarity we use later in modeling the system (§3.1). In §3.2, we investigate the stellar variability, showing that the output has been constant within a few percent and that stellar variations do not drive the changes seen in the infrared. Consequently, those changes are driven by motions of the dust grains produced in planetesimal collisions, as we discuss in §3.3. In §3.4, we identify planets or low-mass brown dwarfs in a V488 Per planetary system as the drivers of those collisions. With the results of these four sections as boundary conditions, we construct a simple, optically-thin model of the debris disk in §3.5; it consists of an outer, cold disk at ~25 - 45 au radius, an inner ring at 0.30 - 0.35 au, and a population of finely divided dust being dragged from this ring into the star by the stellar wind until it reaches the dust sublimation radius at ~.02 au, where the grains are destroyed. The resulting SED, shown in Figure 4, provides a pictorial summary of the section. In §3.6, we show that the level of activity we are witnessing must involve collisions of large asteroid-sized bodies and that it is unlikely to be sustained for much longer than $\sim 1000 - 10,000$ years. §3.7 briefly summarizes the status of the system.

---

[2] https://www.sdss.org/

Table 4. Multiband Photometry of V488 Per[a]

| Band ($\mu$m) | mJy | error | reference | star only |
|---|---|---|---|---|
| B (0.44) | 12.9 | 0.61 | Mermilliod et al. (1987) | – |
| V (0.55) | 29.17 | 1.31 | This Work | – |
| J (1.24) | 63.39 | 1.21 | 2MASS | 64.2 |
| H (1.66) | 66.55 | 1.75 | 2MASS | 65.6 |
| $K_S$ (2.16) | 49.06 | 1.12 | 2MASS | 44.8 |
| W1[b] (3.35) | 42.8 | – | This Work[c] | 23.2 |
| W2[b] (4.6) | 43.3 | – | This Work[c] | 12.4 |
| 8.8 ($\Delta\lambda$ = 0.8 $\mu$m) | 42 | 4 | This Work | 3.5 |
| 10.5 ($\Delta\lambda$ =1.0 $\mu$m) | 52.5 | 9 | This Work | 2.4 |
| W3[b] (11.56) | 43.68 | 1.4[d] | ALLWISE | 2.0 |
| 11.7 ($\Delta\lambda$ = 1.0 $\mu$m) | 40 | 8 | This Work | 1.9 |
| 12.4 ($\Delta\lambda$ = 1.2 $\mu$m) | 52 | 19 | This Work | 1.7 |
| W4[b] (22.09) | 74.14 | 2.35 | ALLWISE | – |
| 70 | 63.3 | 5.9 | This Work | – |
| 70 | 66.45 | 6.66 | Herschel/PACS point source catalog | – |
| 160 | 25.7 | 12.2 | This Work | – |

[a]The tabulated photometry is for bands either known not to vary over our monitoring period or where the cadence does not document the variations clearly.

[b]ALLWISE for W1 - W4 plus NEOWISE for W1 - W2 photometry

[c]Average flux density for all WISE measurements in this band, supplied for context.

[d]3% error assumed

Table 5. IRAC Photometry on Sept. 26, 2006

| Band ($\mu$m) | mJy | error |
|---|---|---|
| 3.54 | 61.4 | 0.9 |
| 4.49 | 57.1 | 0.9 |
| 5.71 | 63.5 | 0.9 |
| 7.84 | 59.8 | 0.9 |

Table 6. Radial Velocities of V488 Per

| v (km s$^{-1}$) | error (km s$^{-1}$) | JD | reference |
|---|---|---|---|
| -0.1 | 0.6 | – | Stauffer et al. (1985) |
| -0.31 | 0.15 | – | Mermilliod et al. (2008) |
| 1.8 | 0.5 | 2450795 | Zuckerman et al. (2012) |
| 0.64 | 0.87 | – | Gaia DR2 |
| -0.059 | 0.1$^a$ | 2457821.6 | Majewski et al. (2017) |
| 0.337 | 0.1$^a$ | 2458097.8 | Majewski et al. (2017) |
| 0.177 | 0.1$^a$ | 2458148.7 | Majewski et al. (2017) |
| -0.087 | 0.1$^a$ | 2458179.6 | Majewski et al. (2017) |

$^a$Errors according to the SDSS documentation https://www.sdss.org/.

### 3.1 Stellar properties of V488 Per

V488 Per is often described as being a K0 type star (e.g. Zuckerman et al. 2012), but this designation appears to arise from photometric, not spectroscopic, arguments (Allain et al. 1996). Its $T_{eff}$ of ~ 4930 K (Balachandran et al. 1988; Randich et al. 1998), 4906 K (LAMOST), or 4977 K (Bai et al. 2019) suggests a spectral type of K2 - K2.5V (Pecaut & Mamajek 2013)[3]. At the age of the $\alpha$ Per cluster, a K2V star will have settled onto the main sequence. We compare the photometry of the star with a Kurucz model for a main sequence 5000K dwarf in Figure 4. The agreement is excellent, indicating that the type designation is appropriate and that the star is very little reddened. Integrating the Kurucz model fitted to the photometric points (see Figure 4), we derive a luminosity of 0.28 ± 0.03 L$_{sun}$, again midway between K2V and K2.5V (Pecaut & Mamajek 2013).The basic properties of V488 Per are very similar to those of eps Eridani, as shown in Table 7. The latter star provides the closest example of a substantial debris disk and is very thoroughly studied. Given the similarity of the two stars, it will be beneficial to argue by analogy in interpreting some of the behavior of V488 Per and its debris system.

---

[3] also see http://www.pas.rochester.edu

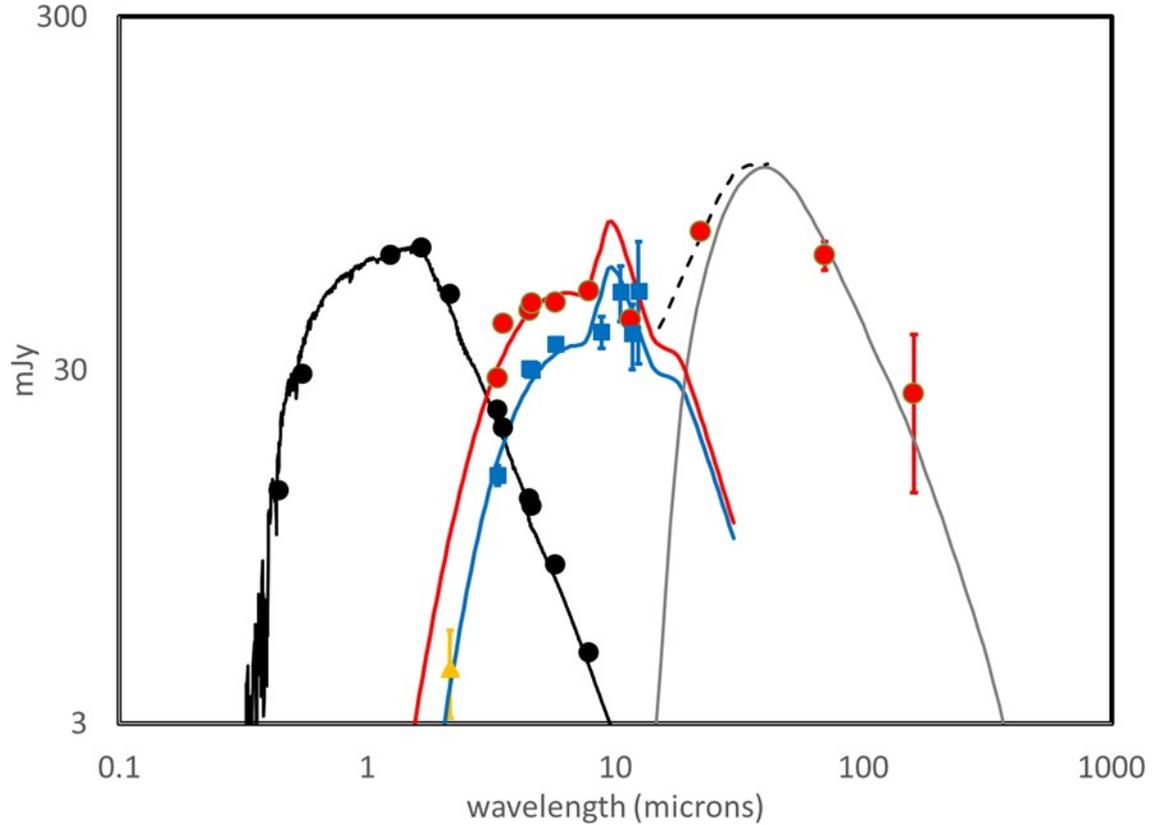

**Figure 4.** Model SED for V488 Per. Photometry in the stellar-dominated regime is shown as black points, which are fitted well by the 5000K Kurucz dwarf star model, as shown. The model is extended to longer wavelengths with the points showing the stellar flux at some of the mid-IR bands. The mid-IR disk measurements have been determined for two output levels (see text), one shown in red and the other in blue with the disk SED model shown for each on the assumption that the amplitude does not affect the silicate emission strength. The disk point at 2.16 $\mu$m is shown in orange because the output level when it was obtained is not known. The variability is not expected to extend into the far infrared, hence only one set of values is shown there (based on a disk extending from 25 to 45 au). The dashed line is the sum of the two disk component fluxes in the range where they are comparable for the red model. These models have 3.6% of the stellar luminosity being reradiated by the outer disk component and 10.1% (blue model) or 16% (red model) by the inner one, for a total fractional excess ($L_{disk}/L_{star}$) of ~ 14 - 20%, reproducing the high fractional luminosity of ~ 16% reported by Zuckerman et al. (2012).

**Table 7.** Comparison of V488 Per with $\epsilon$ Eri

| property | V488 Per | reference | $\epsilon$ Eri | reference |
|---|---|---|---|---|
| spectral type | K2-2.5 V | this work | K2V | Gray et al. (2006) |
| temperature(K) | 4977 | Bai et al. (2019) | 5084 | Kovtyukh et al. (2003) |
| luminosity ($L_0$) | 0.28 ± 0.03 | this work | 0.335 | Bonfanti et al. (2015) |
| rotation period (days) | 5 − 6 | this work | 11.2 | Fröhlich (2007) |
| variability type | BY Dra | Stauffer et al. (1985) | BY Dra | Fröhlich (2007) |
| X-ray luminosity ( ergs s$^{-1}$) | $5 \times 10^{29}$ | Prosser et al. (1996) | $2 \times 10^{28}$ | Coffaro et al. (2020) |
| outer debris ring radius (AU) | ∼ 35 | this work | 64 | Chavez-Dagostino et al. (2016) |
| age (Myr) | ∼ 80 | Soderblom et al. (2014) | ∼ 440 | Barnes (2007) |
| distance (pc) | 173.5 | Gaia EDR3 | 3.22 | Gaia EDR3 |

### 3.1. Optical Variability

Stauffer et al. (1985) reported periodic variations for V488 Per with a peak-to-peak amplitude in $V$-band of ∼ 5% and a period of ∼ 5.15 days. Eleven years later, Allain et al. (1996) found the star to have a peak-to-peak variability amplitude in $V$-band of ∼ 10% with a period of ∼ 6.4 days. In both cases the observations extend only over two periods, but Allain et al. (1996) suggest that the period difference is real and might reflect differential rotation. Heinze et al. (2018) did a Fourier analysis of photometry of the star from the ATLAS survey (ATLAS uses a robotic telescope with a CCD covering 5 degrees and scans half of the accessible sky each night), finding a period of 5.975 days.

We used a phase curve analysis to look for periodic variations in our much longer optical data string, searching for periods between 5 and 6.5 days to bracket the previous reports. There is an indicated periodicity at 5.967 days, in agreement with the value from the ATLAS survey. With this approach, the results are degenerate with a number of indications of different periods, but none is any larger in amplitude than the 5.967 day one. There are also TESS data from November, 2019, for three weeks (i.e., only a few periods). They show a possible period at about 5.24 days, supporting the evidence for differential rotation. The peak to peak amplitude of the variations is ≤ 2%.

BY-Dra-type variations typically only change significantly in amplitude over long timescales - of order a year or more (Alekseev & Kozhevnikova 2018). Therefore, to investigate further, we analyzed just the first year, 2015-2016, which as shown in Figure 5 shows periodic variations with a peak-to-peak amplitude of 5.9%. We then analyzed the second year, then the third, and finally the rest, into early 2020, i.e., including the period when the TESS data were obtained. None of these latter intervals showed evidence for variability with this period greater than 2% peak-to-peak, in agreement with the short sequence obtained with TESS. Figure 6 shows the period plot for all the data past the first year. The difference in behavior between the first year and the remaining time is supported by the rms scatter without removing the variability: 3.1% for the first year, 2.4% for the rest. If we compensate for the variability in the first year, the rms scatter is reduced to 2.1%, indicating that the decrease in rms scatter after the first year is associated with a reduction in the amplitude of the BY Dra type behavior. In addition, the star is about 2% brighter after the first year (see Figure 3), consistent with lower surface coverage by spots (Alekseev & Kozhevnikova 2018). Given that the measurements include contributions from a large number of observers using a similarly large number of instruments, it is likely that a significant contribution to this remaining scatter is systematic errors from one photometric system to another, as well as statistical errors. That is, the variation of the star other than the periodic modulation and possible long term trend, is likely to be < 2%. Although only for a short interval, the TESS data support this conclusion.

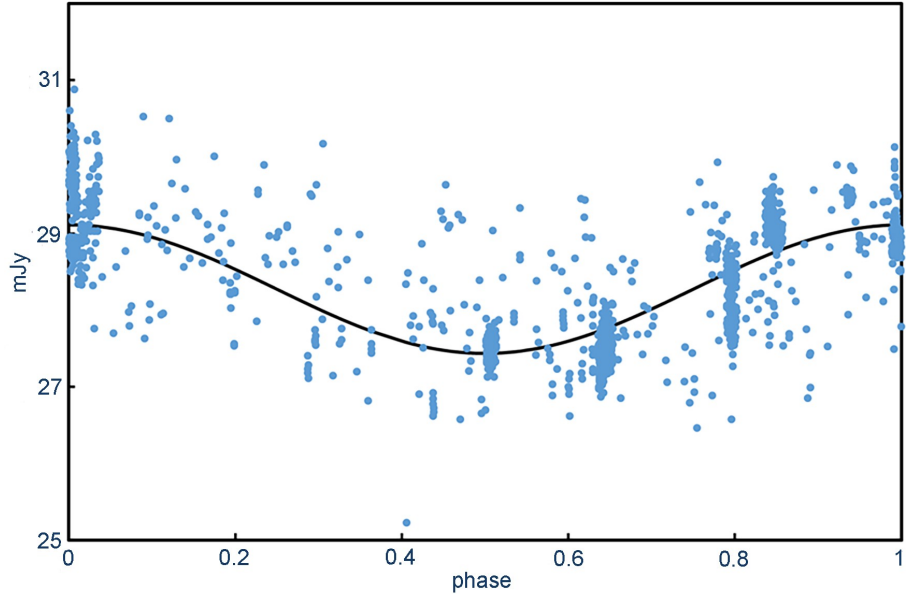

**Figure 5.** Phase plot for the first year of visible photometry of V488 Per. The possible modulation has a period of 5.967 days and the fitted line (black) has a peak-to-peak amplitude of 5.9%.

There is no other evidence for variations in the V-band over the entire set of data; in particular, although flares are sometimes seen in BY Dra variables, there is no convincing evidence in any for this star (other than the single measurement early in our monitoring period). The average magnitude over the 1760 measurements, V = 12.84, agrees with values of 12.83 reported over the past 15 years (e.g., Stauffer et al. 1985; Allain et al. 1996). That is, the star is non-variable to the few % level in V over our entire measurement period, from 2015 to 2020. Even in the first year of our monitoring, where there was a significant amplitude of periodic optical variations, there was no infrared variability at that period. That is, there is no stellar variability that could power the variations seen in the infrared debris disk emission.

### 3.2. *Grain dynamics*

We now move toward modeling the behavior of the debris disk as revealed by the infrared variations. Given a mechanism to stir the planetesimals around V488 Per and cause them to collide, the observational signatures of this process depend on the production of dust in the collisions and their aftermaths, and then the motion of dust grains subsequent to these collisions. The production of copious amounts of dust on a short timescale is common to other extreme and variable debris disks (e.g., Su et al. 2019). However, because of the low stellar luminosity of V488 Per, it is plausible that the subsequent motions of those dust grains differ from the previously studied systems. Here we discuss the grain dynamics as constrained by the variability characteristics.

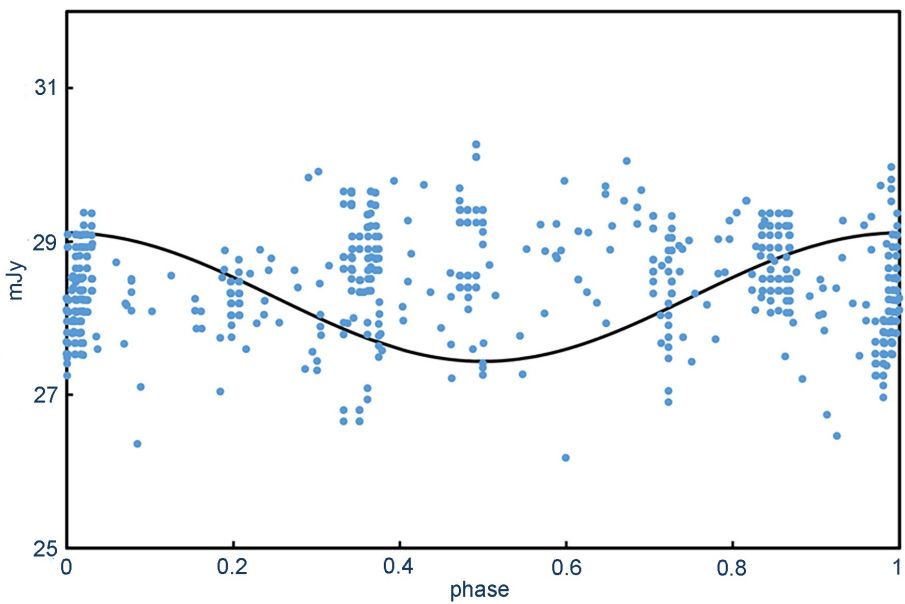

**Figure 6.** Phase plot for the remaining years of visible photometry of V488 Per. The fitted behavior for the first year is shown as the black line. The fit for these data places an upper limit of 2% for variations at the period of 5.967 days for this time interval.

### 3.3.1. *Loss by radiation pressure force*

The infrared light curve (Figures 1 & 2) shows decays after emission peaks with timescales as short as a few months, requiring a rapid grain removal mechanism. Similar behavior is seen in some other variable debris disks such as ID8 (e.g., Su et al. 2019), where the behavior is sampled much better because of the longer visibility window for *Spitzer*. However, in all the other cases, the rapidly variable disks are around more massive and luminous stars. In those situations, once grains are broken down to micron sizes, radiation pressure force blows them outward from the star relative to their birth places. The affected grains will quickly flow to radii where they are at reduced temperatures and hence do not contribute substantially to the infrared signals at 3 - 5 $\mu$m where the variability is observed. Issues remain because a normal collisional cascade will continue to replace the very small grains, but if the grains are produced in ways that avoid this process (e.g., condensed from vapor), then the short timescales are readily explained. However, at the luminosity of V488 Per, first-order estimates, generally assuming solid spherical grains, indicate that radiation pressure force will be too weak to eject grains.

A more complete picture is provided by calculations assuming more realistic grain properties, i.e. irregular shapes, differing compositions. Arnold et al. (2019) report calculations for these more complex cases, which include eps Eri so we can apply their results directly to V488 Per. They show that carbon grains smaller than ~ 1 $\mu$m in size are subject to radiation-pressure blowout, but silicate and ice grains of any size are not (this would be particularly the case at the modestly lower luminosity of V488 Per). Grains of mixed composition with a dominant fraction of carbon were also subject to blowout, but those of predominantly silicates were not. That is, blowout cannot be ignored but at least a large subset of grains are not likely to be removed by this process.

### 3.3.2. *Loss by stellar wind drag*

For grains where radiation pressure force does not overcome gravity, grain dynamics is determined by Poynting-Robertson (*P - R*) drag and the effects of the stellar wind. *P - R* drag is much too slow to account for the variability. However, stellar wind drag has already been shown to be of potential interest for eps Eri (Reidemeister et al. 2011). The strength of the wind is proportional to X-ray surface flux (Wood et al. 2002), which has been measured to be 25 times higher for V488 Per than for eps Eri (see Table 7). Although this conclusion rests on a single measurement obtained 25 years ago, the fact that the X-ray luminosity of V488 Per was found to be typical for the K-type members of the *a* Per cluster, of which 83% were detected, indicates that it should be indicative of the current X-ray output. Thus, although the wind does not appear to dominate the debris disk properties for eps Eri (Su et al. 2017), it seems likely that the stellar wind plays a central role in the grain dynamics around V488 Per.

The particles in a stellar wind (i.e., mostly energetic protons) impact dust grains and they recoil anisotropically. The process is analogous to *P - R* drag. Plavchan et al. (2005) utilize this analogy to provide an estimate of the ratio of the relevant timescales:

$$\frac{\tau_{P-R}}{\tau_{SW}} = \frac{Q_{SW}}{Q_{P-R}} \frac{c^2 \dot{M}_{SW}}{L_*} \qquad (1)$$

where $Q_{SW}$ and $Q_{P-R}$ are the coupling coefficients and $\dot{M}_{SW}$ is the mass loss in the wind. For $Q_{SW}/Q_{P-R} = 1$ and $\dot{M}_{SW} = 2 \times 10^{-14} M_{sun}$ yr$^{-1}$ (values appropriate for the sun), the ratio is 0.37. $\dot{M}$ for eps Eri is about 30 times that for the sun (Wood et al. 2002), and taking $\dot{M}_{SW}$ for V488 Per to be 25 times higher still, we get $\tau_{P-R}/\tau_{SW} \sim 300$ for V488 Per. From Gustafson (1994), the time required for a particle under $P - R$ drag to fall into a star can be estimated to be

$$\tau_{P-R} \approx \frac{400}{\beta} \frac{M_*}{M_\odot} \left(\frac{r_0}{au}\right)^2 \, yr, \qquad (2)$$

where $\beta$ is the ratio of radiation pressure to gravitational force and $r_0$ is the starting distance from the star. Since the sublimation radius for plausible grain properties is 0.01 - 0.02 au, this equation is also appropriate for a grain to reach this radius. For grains between 0.1 and 1 $\mu$m in size, we can take $\beta = 0.2 - 0.4$ from Arnold et al. (2019). Equation (2) gives a time of ~ 300 - 150 yrs for such a grain to be destroyed by falling from the inner debris ring at 0.3 au (see next section) to the sublimation radius through *P - R* drag, confirming that this process is much too slow to account for the variability. However, these estimates convert to only 6 - 12 months for stellar wind drag to bring a grain inwards from 0.3 au. This is a sufficiently good match to the observed behavior to identify stellar wind as the driving effect causing the grains to spiral inward and be lost by sublimation, leading to a drop in the debris disk flux after a peak.

This loss mechanism can also account for the increase in color temperature between 3.6 and 4.5 $\mu$m when the flux in these bands increases (see Figure 1). If an increase marks a major collision, grains of size 2 $\mu$m will migrate rapidly inward where they are heated toward their sublimation temperatures. Grains larger than 2 $\mu$m have smaller values of $\beta$ (Arnold et al. 2019) and will remain closer to their creation positions, where they can participate in collisional cascades to restore small grains to maintain their equilibrium value in the disk. That is, a qualitative description of the variations is that there is a belt of planetesimals with a highly elevated rate of collisions. When a particularly large collision occurs, we see an elevation of the debris disk reradiated energy, which we have been able to monitor between 3 and 5 $\mu$m but presumably also affects the longer wavelengths, e.g., 10 $\mu$m. This behavior decays and the dust grains are heated and destroyed as they spiral into the star under the influence of stellar wind drag.

### 3.3. *What drives the variability?*

At the age of $\alpha$ Per (80 Myr), the most durable protoplanetary disks will have dissipated (Meng et al. 2017), requiring that the V488 Per disk originate as a massive debris disk, i.e., replenished by planetesimal collisions rather than being primordial. The vast majority of debris disks at this age have settled into a regime with fractional excesses $\ll 1\%$ and are geometrically thin. The high fractional excess of the inner ring in the V488 Per disk requires not just that it be very dense, but that it have sufficient thickness out of the disk plane to capture a large fraction (i.e., 10 15%) of the stellar luminosity, requiring a thickness of order 0.1 au. Along with the variability, these attributes require that the disk must undergo substantial disturbances that stir it to enhance the collisional rates. We consider three possibilities: (1) coronal activity; (2) stirring by nearby stars; or (3) stirring by members of a V488 Per planetary system.

#### 3.3.1. *Coronal Activity*

It has been suggested that rapid debris disk variability is triggered by coronal ejections and similar activity from the star (Osten et al. 2013). The overall X-ray luminosity that indicates a typical level of coronal activity rather than an extraordinary one, plus the behavior of the BY Dra variability, with a low amplitude during most of the period when we observe substantial infrared variations, would seem to contradict this hypothesis. Other than an initial possible flare, there is no indication in the optical for flare-like events. In addition to the lack of evidence for extraordinary coronal activity, it is hard to understand how this mechanism can produce major increases in the amount of dust, particularly as seen at the end of our Spitzer monitoring. We therefore discard the possibility of the infrared variations being related directly to coronal activity.

#### 3.3.2. *Possible stellar companions*

Zuckerman (2015) has suggested that the extreme debris disk phenomenon may be linked to close stellar companions and the resulting possible perturbations. Specifically for V488 Per, he identified two X-ray-emitting stars with proper motions consistent with membership in the $\alpha$ Per cluster as possibly being linked with the huge infrared excess. One of these stars, APX 43A, has a parallax from Gaia EDR3 of 5.914 ± 0.013 mas, consistent with cluster membership but is indicated to be 4.4 ± 0.6 pc in front of V488 Per. The second star, APX 43B, has a Gaia EDR3 parallax of 7.081 ± 0:27 mas compared with 5.764 ± 0.012 mas for V488 Per and thus appears to be roughly 40 pc in the foreground of the cluster. However, as already indicated by the relatively large quoted error, the fit to this star is poor and its parallax may be consistent with cluster membership[4]. The projected distance of this star from V488 Per is 70 au, but the debris system model discussed below indicates that the physical separation is larger. The presence of a cold debris disk component at a nominal radius of 25 - 45 au (as discussed in the next section) is probably incompatible with a stellar mass companion closer than $\sim 100 - 135$ au (Yelverton et al. 2019).

### 3.5 *Simple Model*

To obtain a rough idea of the layout of the circumstellar material around V488 Per, we have computed an optically thin debris disk model[5]. This fit suggests the need for three disk components: (1) an outer ring at 25 45 au that dominates the far infrared; (2) a ring at 0.30 ⊖ 0.35 au; and (3) a distribution of micron-sized grains extending from 0.3 au inward to the sublimation radius and representing the grains being dragged inward by the stellar wind. A flat distribution with radius is appropriate for grains migrating under stellar wind drag (e.g., van Lieshout et al. 2014). The $r^2$ dependence of the inflow timescale in equation (2) already implies this behavior.

The modeling begins by assembling two spectral energy distributions, corresponding to the two times where we can assemble the appropriate measurements across the 3 - 15 $\mu$m range and representing two disk emission levels. Figure 2 illustrates the assembly of the data for the models. The first period, which uses the cryogenic WISE data from BMJD 55243 (15 Feb. 2010, the first WISE measurements in Figure 2), is shown in red dots in Figure 4. We combine the WISE data with the full IRAC set in Table 5; at least for the two bands at almost the same wavelength (IRAC2 and W2), the flux levels are similar. The shorter wavelength IRAC band indicates a higher temperature than is typical, resulting in a point that lies somewhat above our fitted continuum. We include the W4 and Herschel data in both SEDs, since we will show that it is unlikely that the variability extends to these wavelengths. The second model will apply for the times when we obtained the COMICS observations. To construct a full SED requires that we relate these measurements to the level of activity in the shorter bands. Fortunately, Figure 1 shows that the variability over any 45-day interval is modest compared with the full amplitude. The second set of COMICS measurements (28 Jan 2018, BMJD 58147) was followed in three days by a set of WISE W1 and W2 photometry, with a value of 41.2 mJy for W2. The first set of COMICS measurements (14 Jan 2017, BMJD 57768) was preceded by 11 days with IRAC2 data and followed by 24 days with a W2 measurement; the flux density dropped by only 9% between them. We estimated a value at the time of the COMICS data of 43.8 mJy in W2 by linear interpolation. Since the two sets of COMICS measurements were obtained when the shorter wavelength activity was at closely the same level, we averaged the results from the two runs. Figure 4 also shows the excess in $K_S$-band from the 2MASS photometry; since this cannot be associated with any particular level of activity it is plotted as an orange triangle.

---

[4] The indicated error is ∼ 2.2 mas from the quality of the fit
[5] Using Debris Disk Simulator (Wolf & Hillenbrand 2005)

The model of the outer disk is appropriate for both activity levels. We fitted it with a dust mass of 0.05 $M_\oplus$ in grains between 0.1 $\mu$m and 1 mm in size with a power law grain size index of 3.65, distributed with grain density falling as $r^{-1.5}$ where $r$ is the distance from the star, and placed at 25 - 45 au. The total mass including bodies > 1 mm in size would be much larger. We assumed grains with the optical constants derived as providing the best fit to the outer disk of $\beta$ Pic by Ballering et al. (2016).

For the inner material, we start by modeling the blue points. We first explored possible grain compositions; the modeling of Ballering et al. (2016) of the outer $\beta$ Pic disk need not apply to dust close to a star. In fact, we found that overall fits to the spectrum at wavelengths < 8 $\mu$m could be obtained with dust of that composition or with astronomical silicate, placed at 0.13 - 0.15 au from the star with a radius-independent distribution inward to the sublimation radius. However, these fits (and ones using crystalline silicates in general) produced a very strong silicate emission feature radiated by the small grains and this meant the fit to the data at 8 -13 $\mu$m was poor. We therefore took the inner ring to be 50% astronomical silicate and 50% carbon. We found a good fit with a mass of $6 \times 10^{-6}$ M with lower and upper grain size limits of 0.1 $\mu$m and 1 mm; this component was placed at 0.3 - 0.35 au from the star with the same grain size slope and density behavior relative to the star as for the first component.

For the third component, we assumed the same grain composition but included grains from 0.03 to 2 $\mu$m in size and distributed them with constant density from the inner radius of this ring, 0.3 au, to the sublimation radius at 0.01 au (we took the sublimation temperature of the carbon to be 2000 K, given the short duration of the exposure of the grains to the highest temperatures). The required grain mass is $2.5 \times 10^{-8}$ M . The overall fit, illustrated in Figure 4, is reasonable with this model. In it, 80% of the luminosity from the inner zone (r < 20 au) is radiated by the ring and 20% from the inner component representing the small grains being dragged into the star.

The red points were modeled by assuming an increase in the inner, dragged-in component with no other changes. A good fit was obtained (see Figure 4) by increasing the dragged-in dust output by a factor of 4.5, so they contribute about half of the luminosity of this inner dust component in this case. These values are compatible with the change in fractional luminosity derived below.

The model indicates that 3.6% of the stellar luminosity is reradiated by the outer disk component and 10.1% (blue model) or 16% (red model) by the inner one, for a total fractional excess ($L_{disk}/L_{star}$) of ~ 14 - 20%. In addition, the color temperature based on synthetic photometry to reproduce the IRAC1 and IRAC2 bands for the red points is about 80K higher than with the blue point model, in agreement with the observations in Figure 1. In addition to the extreme level of activity in the inner component, the 3% fractional luminosity of the outer disk component is also very high, indicating that the entire disk system is in a heightened state of activity. However, in the model the outer disk component does not partake in the variability.

In general, models of this nature for debris disks are notoriously degenerate, and the limited constraints and unique properties for this one make this even more the case. Our goal, therefore, is not to generate a unique model, but to show that the behavior can be explained within the constraints. However, the general properties appear to be robust. For example, fitting the far infrared points with a blackbody model and applying a correction as in equation 8 of Pawellek & Krivov (2015) indicates that the outer ring should have a radius of 35 au, i.e., similar to the result above. The rapid variations of the 3 - 5 $\mu$m flux indicate the need for the inner component, since it is required to produce the necessary high grain temperatures. It is not possible to match the flux density at 70 $\mu$m with this component and any plausible grain properties, thus requiring the outer component. That is, although more sophisticated models, e.g. including optically thick ring components, will modify the parameters, they are unlikely to avoid the necessity for two dust rings with the variations originating in one around 0.3 au from the star. The possibility of a lower density of grains between these components is not excluded - and in fact should exist due to stellar wind drag on grains in the outer component.

### 3.4. *Required Masses*

We can estimate the mass of the colliding objects that produce the dust as follows. We consider the major event in 2019 and the increase in dust of size between 0.3 and 2.0 $\mu$m that it requires over the more quiescent level of activity (from Figure 2, we take this to be twice the increase of the model for the red points over that for the blue ones). The resulting mass applies to grains between 0.3 and 2 $\mu$m in size; integrating up to a meter in size, and assuming a density of 3 g cm$^{-3}$, we estimate that the amount of material in the collisional cascade to produce the observed dust is equivalent to a single object of radius 85 km, or if divided between two equal-sized bodies, each would be 60 km in radius. That is, this event alone represents the collision and breakup of two bodies similar in size to large asteroids in the Solar System.

Another consequence of the model is that the dust mass flow into the star is high, of order $10^{17}$ - $10^{18}$ kg yr$^{-1}$ depending on whether the infall time is 1 or 0.1 years. This is equivalent to the mass of Ceres every 10,000 - 1000 years. That is, the extreme level of debris disk activity around V488 Per must be very shortlived. Given that it is by a considerable margin the most extreme debris disk known, this conclusion is perhaps expected. At the age of 80 Myr, models suggested that the system should be past the era of the largest planetesimal impacts (e.g., Genda et al. 2015). Therefore, the extremely large amount of debris suggests that the planetary system around the star (as discussed in Section 3.4.3) is undergoing a major readjustment, perhaps roughly analogous to that resulting in the Late Heavy Bombardment in the Solar System.

### 3.5. *Summary*

The infrared excess of V488 Per is not only extremely large but is also variable on timescales of a few months. This behavior has persisted for at least 15 years, but the mass required to sustain it suggests that it can have persisted for no more than 1000 - 10,000 years. We interpret our results in terms of a massive planet (or low mass brown dwarf) perturbing the members of an asteroid-like belt about 0.3 au from the star, resulting in a high level of planetesimal collisions. The dust that results (largely from secondary collisions) migrates inward toward the star due to drag with the stellar wind. V488 Per is 80 Myr old, past the primary period for high rates of planetesimal collisions as terrestrial planets form in classical simulations of terrestrial planet formation (Genda et al. 2015). However, the collisional activity in its planetesimal system is perhaps the most extreme known. As a result, although an extreme collision rate associated with classical terrestrial planet formation scenarios is still possible (Chambers 2013), we may be witnessing a delayed instability as suggested in some recent simulations (e.g., Clement et al. 2020), i.e., with the high rate of dust production arising from migration or other forms of orbital instability within its (unseen) system of massive planets.

## 4. CONCLUSION

We report on a detailed study of the variations in the infrared emission by the most extreme known debris disk, that around the 80 Myr-old K2-2.5V star, V488 Persei. Our observations and analysis of them show that:

- The debris disk emission varies at a very high level, comparable to the most variable systems known (e.g., ID8 in NGC 2547).

- The star itself shows only small variations of the BY Dra variety, and does not drive the infrared variations.

- The debris system consists of an inner and an outer component, both of which have very high fractional luminosity (~ 3.6% for the outer component, 10 - 16% or even higher for the inner one).

- The variations in the output of the inner component are most likely initiated by gravitational stirring of a belt of planetesimals at 0.3 au by a low-mass brown dwarf or a planet in the V488 Per system. This stirring is resulting in a high rate of collisions among massive planetesimals, resulting in episodic outbursts of dust production.

- The dust grains generated as a consequence of this stirring migrate inward toward the star due to stellar wind drag, and eventually sublimate at about 0.01 au from the star. As a result, they heat as they leave their place of formation.

- Evidence for this type of migration is the increase in color temperature when the excess is bright

- This behavior contrasts with that of ID8, where the dust color temperature remains constant, consistent with the grains in that system being blown *away* from the star by radiation pressure force, so they cool as they leave their place of formation.

- The mass loss rate from the V488 Per inner debris ring is extremely high and it is being produced by collisions of large planetesimals, indicating that the extreme excess and fractional luminosity will be short-lived. There is truly an exceptional level of collisional activity in the planetesimals around this star.

- This extreme level of collisional activity is occurring relatively late in the system evolution. Although it might be a late event in the process of terrestrial planet formation, it may also signal a re-adjustment in the star's planetary system roughly analogous to the Late Heavy Bombardment in the Solar System.

The behavior of V488 Per along with that of ID8 in NGC 2547 and other related variable debris disks establishes that short-lived extreme episodes of planetesimal collisions are fairly common during the first 100 Myr of planetary system evolution. However, the details of this behavior for this star are unique in showing evidence that the mechanism for the loss of finely divided dust is stellar wind drag, rather than radiation pressure force as is typical of more massive and luminous stars. The system also supports the possibility that some of the extreme systems around relatively old ($\gtrsim$ 50 Myr) stars are experiencing major readjustments of their planetary systems, rather than representing a continuation of the early phase of terrestrial planet building represented in classical simulations.

## 5. ACKNOWLEDGEMENTS

This work has been supported by NASA Exoplanets Research Program Grant 80NSSC20K0268, NASA Grants NNX13AD82G, NNX17AF03G and 80NSSC20K1002. C.M. acknowledges support from NASA ADAP grant 18-ADAP18-0233. We thank Schuyler Wolff for helpful comments, and the referee for a speedy and helpful report. The paper is based on observations made with the Spitzer Space Telescope, which was operated by the Jet Propulsion Laboratory, California Institute of Technology. This publication also makes use of data products from the NearEarth Object Wide-field Infrared Survey Explorer (WISE), which is a project of the Jet Propulsion Laboratory/California Institute of Technology. WISE is funded by the National Aeronautics and Space Administration. We acknowledge with thanks the variable star observations from the AAVSO International Database and the infrastructure maintained by AAVSO, which were used in this research. We thank Peter Milne for obtaining additional visible photometry; likewise, we have used photometry from the ASAS-SN project, for which we are grateful. The paper also utilizes data collected by the TESS mission. Funding for the TESS mission is provided by the NASA's Science Mission Directorate. This work also includes results from the European Space Agency (ESA) space mission Gaia. Gaia data are being processed by the Gaia Data Processing and Analysis Consortium (DPAC). Funding for the DPAC is provided by national institutions, in particular the institutions participating in the Gaia MultiLateral Agreement (MLA). The Gaia mission website is https://www.cosmos.esa.int/gaia. The Gaia archive website is https://archives.esac.esa.int/gaia. We also include data from the *Herschel* mission. *Herschel* is an ESA space observatory with science instruments provided by European-led Principal Investigator consortia and with important participation from NASA. Finally, we used data from the Two Micron All Sky Survey (2MASS), a joint project of the University of Massachusetts and the Infrared Processing and Analysis Center/California Institute of Technology, funded by the National Aeronautics and Space Administration and the National Science Foundation.

*Facility:* AAVSO, Spitzer, Herschel, TESS, Gaia, WISE, 2MASS, Subaru